# Gated Spin Transport through an Individual Single Wall Carbon Nanotube

*Bhaskar Nagabhirava,[1]Tanesh Bansal, Gamini Sumanasekera,[2]Lei Liu,[3]and Bruce W. Alphenaar[1]**

[1]Department of Electrical and Computer Engineering, University of Louisville, Louisville, KY, 40292

[2]Department of Physics, University of Louisville, Louisville, KY, 40292

[3]Department of Physics, McGill University, Montreal, Quebec, Canada, H3A 2T8

*Corresponding Author: Phone: 502-852-1554; Fax: 502-852-1577; Email: brucea@louisville.edu

ABSTRACT   Hysteretic switching in the magnetoresistance of short-channel, ferromagnetically contacted individual single wall carbon nanotubes is observed, providing strong evidence for nanotube spin transport.  By varying the voltage on a capacitively coupled gate, the magnetoresistance can be reproducibly modified between +10% and -15%.  The results are explained in terms of wave vector matching of the spin polarized electron states at the ferromagnetic / nanotube interfaces.

Due to their unique structural and electrical properties, carbon nanotubes have been investigated extensively for possible electronic device applications.[1] Carbon nanotube *spin* electronics, while less thoroughly explored, also holds substantial promise.[2] Carbon nanotubes have large electron scattering lengths (measured to be 10 microns or higher at room temperature)[3] and weak spin orbit coupling, so that the nanotube spin scattering length is expected to be extremely large. Initial evidence for spin transport through nanotubes was provided by the observation of hysteretic magnetoresistance switching in ferromagnetically contacted multi wall nanotubes (MWNTs).[4] These results have since been reproduced and expanded on by a number of authors.[5,6]

There have also been numerous attempts to observe evidence for spin transport through ferromagnetically contacted single wall nanotubes (SWNTs).[2, 7-10] SWNTs offer many advantages over MWNTs for spin transport studies, including increased scattering lengths, well-defined electronic band structure, enhanced Coulombic interactions (leading to novel physical phenomena), and the possibility to modify the nanotube resistance with a capacitively coupled gate. A less obvious advantage of the SWNT is that the SWNT resistance shows a much weaker dependence on magnetic field than the MWNT.[11] The large intrinsic MWNT magnetoresistance in combination with the fringing field from the ferromagnetic contacts can produce resistance changes that mimic those due to spin transport.[2] Despite these advantages, measurements of SWNTs have so far provided less than ideal evidence for spin transport. Some groups have reported magnetoresistance switching in ferromagnetically contacted SWNTs,[2,7,8] while others fail to observe even a background magnetoresistance.[9] Four-terminal measurements of SWNTs provide evidence that some fraction of the observed magnetoresistance is due to spin transport through the nanotubes.[10] However, magnetoresistance switching has recently been reported in SWNT devices having only one semiconducting ferromagnetic contact, which brings into question the validity of interpreting any of the magnetoresistance data in terms of spin transport.[8]

There is no clear answer for why it has been so difficult to demonstrate SWNT spin transport in a reliable manner. One possibility though, is that the nanotube transport length must be made much shorter than simple scattering considerations would imply in order for spin mediated resistance changes to be observed consistently. Experiments suggest that at low temperatures, carbon nanotubes can behave as one or a series of conducting islands (rather than as a ballistic wire).[12] Transport through such small islands will be dominated by Coulomb charging effects, which will greatly increase the electron transport time, and consequently provide more time for spin scattering to occur.[13] Another factor that has not been fully appreciated is the influence of Lüttinger liquid phenomena,[14] which could provide additional pathways for spin scattering that are not available in a Fermi liquid. Irrespective of the exact spin scattering mechanism, it is reasonable to assume that reduction of the nanotube transport length should improve the chance of observing spin transport. Once reliably observed, it will then be possible to characterize the spin scattering mechanisms and optimize the spin mediated resistance signal.

We have fabricated and characterized ferromagnetically contacted "short channel" SWNT devices that show clear hysteretic switching in the magnetoresistance, and provide strong evidence for SWNT spin transport. The main difference between our work and previous studies is that we have greatly reduced the transport length separating the ferromagnetic contacts to distances on the order of 10 nm. Preliminary measurements demonstrate this reduction to be extremely beneficial. We have observed clear hysteretic switching in the magnetoresistance in 75% of our devices, and are able to modify the magnetoresistance between +15% and -10% as a function of gate voltage. The gate mediated change in magnitude *and sign* of the magnetoresistance switching allows us to discount other non-spin related sources for the observed signal and provides the basis for the first SWNT spin transistor.

Our fabrication procedure is outlined in Fig. 1. A grid of high-resolution alignment marks is defined on the surface of an oxidized silicon wafer using e-beam lithography and wet etching. The wafer is then dipped into an iron(III)nitride catalyst solution, and single wall nanotubes are grown on the catalyzed surface by placing the wafer in a $CH_4$ atmosphere at $910^{\circ}$C for 3 minutes. This produces well-separated individual SWNTs that are 1-2 nm in diameter, and approximately 15 microns long. After growth, the SWNTs are mapped out with respect to the alignment grid using an atomic force microscope. A two-step process is then performed to define the nanotube contacts. First, the left contact is defined using standard e-beam lithography and e-beam evaporation (Fig. 1(a)). Next, the right contact is defined in a similar fashion, however, during the evaporation step, the sample is angled so that the line of sight to the right contact is partially shadowed by the left contact (Fig. 1(b)). Using this procedure, the contact separation can be made arbitrarily small, simply by varying the deposition angle. Figure 1(c) shows an SEM image of a finished device, having nickel contacts. The contact separation distance is approximately 10 nm. For three-terminal operation, the ferromagnetic contacts form the device source and drain, while the silicon substrate (which is heavily doped) acts as the gate terminal.

Fig 2(a) shows the two-terminal magnetoresistance of the device shown in Fig. 1(c), measured at 4.2 K, and with zero voltage on the gate. Measurements were made using a standard lock-in detection scheme using a 100 µV excitation voltage with magnetic field from a superconducting magnet directed parallel to the contacts, and perpendicular to the current flow. (The existence of an appreciable conductance at cryogenic temperatures indicates that this is a metallic nanotube.) As would be expected for spin transport through the nanotube, the resistance is high near zero field when the ferromagnetic moments in the two contacts are anti-parallel, and low when the ferromagnetic moments are parallel.[15] The percent change in resistance $\Delta R / R = 2 \left( R_a - R_p \right) / \left( R_a + R_p \right)$ is approximately 10%, where $R_a$ and $R_p$ are the resistances in the anti-parallel and parallel configurations, respectively. Results from other similar devices demonstrate that the short-channel contacting scheme produces a greatly improved yield compared with the standard fabrication technique. Three out of a batch of four devices that were measurable at low temperature showed substantial change in the resistance as a function of magnetic field at 4.2 K (see Figs 2(b) and (c)). By comparison, only one out of more than 100 devices that we measured having contact separations of 100 nm or more showed any sign of magnetoresistance switching. (Recently published work on Fe contacted SWNT devices reports a yield of 4 out of 30 for long channel devices.)[8]

Surprisingly, the magnetoresistance trace in Fig. 2 (c) shows that the resistance is minimized in the anti-parallel configuration (near zero field) and maximized in the parallel configuration (above 100 mT) opposite to the predictions of the simple Julliere model.[15]  All three of the measurements in Fig. 2 were performed with zero bias on the gate.  Further measurements show that the gate bias can be used to alter the behavior of the device magnetoresistance between the standard switching ($R_a > R_p$) and anomalous switching ($R_p > R_a$) states.  Figure 3(a)-(c) shows the magnetoresistance ratio of the device in Fig 2 (b) as a function of magnetic field directed parallel to the contacts for three different values of the gate bias. $\Delta R/R$ is clearly dependent on gate bias, varying from approximately +10% to -6% as the bias changes from 1.44 to 2.76 V.  In addition, as shown in Fig. 3(b), at certain values of gate bias, very little magnetoresistance is observed.  To further elucidate this behavior, we performed a set of similar magnetoresistance measurements on this device at 250 equally spaced values of gate bias between -3V and 11V  The results of these measurements are compiled in Fig. 4, where $\Delta R/R$ is plotted as a function of gate bias for (a) negative and (b) positive sweep directions. While somewhat noisy, a series of reproducible fluctuations can clearly be observed.  Regimes are observed in which $\Delta R/R$ is positive, negative, or close to zero.

Because the contact separation is so small, it is important to estimate the magnitude of the current due to leakage via tunneling through the oxide separating the two contacts. Figure 4(c) shows the conductance of the nanotube device as a function of bias on the silicon substrate at B = 100 mT, where the contact magnetizations are in the aligned, or parallel configuration.  (These data points were extracted from the same 250 magnetoresistance measurements summarized in Figs. 4(a) and (b).)  The conductance is observed to fluctuate by approximately 800% as a function of gate bias.  These fluctuations are most likely due to the combined influence of Coulomb charging and quantum coherence on the transmission through the nanotube. Our results demonstrate that the primary contribution to the conductance is transport through the nanotube and not leakage, since the leakage current should be independent of gate voltage. The maximum possible conductance due to leakage is equal to the minimum device conductance or approximately $5 \times 10^{-8}$ mhos (corresponding to a resistance of $2 \times 10^{7}$ $\Omega$). A simple circuit analysis shows that this amount of leakage is insufficient to account for the magnitude of the observed magnetoresistance switching, and is clearly unable to explain the change in sign in the magnetoresistance as a function of gate voltage.

Since characterization of the short-channel device requires two-terminal measurements, we must also consider the influence that fringing fields from the ferromagnetic contacts have on the intrinsic SWNT magnetoresistance. For two important reasons, however, we feel that this effect is not a significant source of magnetoresistance in our measurements. First, recent results have shown that, in contrast with MWNTs, the magnetoresistance of SWNTs is very small, and amounts to less than 1% at a field of 1 T at 4.2K.[11] This is confirmed by our own measurements, which show no appreciable magnetoresistance beyond the switching that we observe at low fields. Second, this effect does not explain the dramatic improvement in yield that we have achieved simply by decreasing the contact separation. Fringing field effects should be equally important at large and small contact separations. For similar reasons, we feel that our results cannot be related to the magnetoresistance switching recently observed in SWNTs with single ferromagnetic semiconducting contacts.[8] Any single contact effect (whose origin is yet to be explained) should be observed equally for large and small contact separations.

We can qualitatively understand the mechanism behind the gate dependent magnetoresistance in terms of a simple one-dimensional model that takes into account wave function matching across the ferromagnetic / nanotube / ferromagnetic junction, with different Fermi wave vectors for the spin-up and spin-down electrons in the ferromagnetic regions[16] and a variable Fermi wave vector of the nanotube tuned by gate voltage.[17] Similar models have previously been applied to magnetic tunnel junctions[18] and to MWNTs.[6] The electron transmission through the junction for the parallel ($T_p$) and antiparallel ($T_a$) configurations is calculated by solving the one-dimensional Schrodinger equation. The resistance ratio is then approximated by $R/R = 2(T_p-T_a)/(T_a+T_p)$. Because of the different dispersion and different Fermi wave vectors for the two spin states and the difference of the Fermi wave vectors in the ferromagnetic regions and nanotube region, the outcome of the matching, and hence the transmission coefficient depends on both the magnetic configuration and gate bias. As the gate bias increases, the transmission coefficients $T_a$ and $T_p$ oscillate, each out of phase from one another, causing oscillation of $R/R$, as shown in the figure. A more precise description will most likely require treating the nanotube as a quantum dot, and include the influence of Coulomb charging and the zero dimensional electron energy spectrum.[19] However, the simple model does indicate that oscillations in the magnitude and sign of the magnetoresistance are expected provided that the nanotube wave vector can be adjusted independently of the contacts.

In summary, we have fabricated and characterized ferromagnetically contacted SWNT devices with extremely short contact separation distances. The short-channel allows for the observation of magnetoresistance switching whose gate dependence provides strong evidence for spin transport through the SWNT. We note that the short channel contacting scheme is generally applicable to non-ferromagnetic contacts as well, and provides a straightforward technique for fabricating SWNT quantum dot devices.

ACKNOWLEDGMENT The authors thank R.W. Cohn, Q. Si and J. Kono for valuable discussions. Funding provided by ONR / NSF (No. ECS-0224114) and NASA (No. NCC 5-571).

REFERENCES

(1)   Dresselhaus, M.; Dresselhaus, G.; Avouris, P. (Eds.), *Carbon Nanotubes: Synthesis, Structure, Properties and Applications*, Springer: Berlin, 2001.

(2)   Alphenaar, B.W.; Chakraborty, S.; Tsukagoshi, K. In *Electron Transport in Quantum Dots*; Bird, J.P., Ed.;  Kluwer Academic / Plenum Publishers; New York, 2003; pp. 433-456.

(3)   Bachtold, A.; Fuhrer, M.S.; Plyasunov, S.; Forero, M.; Anderson, E.H.; Zettl, A.; McEuen, P.L. *Phys. Rev. Lett.* **2000**, *84*, 6082; Yao, Z.; Kane, C.L.; Dekker, C. *Phys. Rev. Lett.* **2000**, *84*, 2941.

(4)   Tsukagoshi, K.; Alphenaar, B.W.; Ago, H. *Nature* **1999**, *401*, 572.

(5)   Orgassa, D.; Mankey, G.J.; Fujiwara, H. *Nanotechnology* **2001**, *12*, 281; Alphenaar, B.W.; Tsukagoshi, K.; Wagner, M. *J. Appl. Phys.* **2001**, *89*, 6863; Zhao, B.; Monch, I.; Muhl, T.; Vinzelberg, H.; Schneider, T.M. *J. Appl. Phys.* **2002**, *91*, 7026; Sahoo, S.; Kontos, T.; Schonenberger, C.; Surgers, C. *Appl. Phys. Lett.* **2005**, *86*, 112109.

(6)   Chakraborty, S.; Walsh, K.; Alphenaar, B.W.; Liu, L.; Tsukagoshi, K. *Appl. Phys. Lett.* **2003**, *83*, 1008.

(7)   Kim, J.R.; So, H.M.; Kim, J.J.; Kim, J. *Phys. Rev. B* **2002**, *66*, 233401.

(8)   Jensen, J.; Hauptmann, J.R.; Nygard, J.; Lindelof, P.E. Phys. Rev. B **2005**, *72*, 035419.

(9)   Chen, Y.F.; Fuhrer, M.S.; Chung, S.H.; Gomez, R.D. *Bulletin of the American Physical Society*, **2003**, *48*, 1009.

(10)  Tombros, N.; van der Molen, S.J.; van Wees, B.J. Preprint, *cond-mat/* 0506538.

(11) Sagnes, M.; Raquet, B.; Lassagne, B.; Broto, J.M.; Flahaut, E.; Laurent, C.; Ondarcuhu, T.; Carcenac, F.; Vieu, C. *Chem. Phys. Lett.* **2003**, *372*, 733.

(12) Bockrath, M.; Cobden, D.H.; McEuen, P.L.; Chopra, N.G.; Zettl, A.; Thess, A; Smalley, R.E. *Science* **1997**, *275*, 1922.

(13) Brataas A.; Wang, X.H. *Phys. Rev. B* **2001**, *64*, 104434.

(14) Kane, C.; Balents, L.; Fischer, M.P.A. *Phys. Rev. Lett.* **1997**, *79*, 5086; Egger R.; Gogolin, A.O.; *Phys. Rev. Lett.* **1997**, *79*, 5082; Balents, L.; Egger, R. *Phys. Rev. Lett.* **2000**, *85*, 3464.

(15) Julliere, M. *Phys. Lett.* **1975**, 54A, 225.

(16) Stearns, M.B. *J. Magn. Magn. Mater.* **1977**, *5*, 167.

(17) Lemay, S.G.; Janssen, J.W.; van den Hout, M.; Mooij, M.; Bronikowski, M.J.; Willis, P.A.; Smalley, R.E.; Kouwenhoven , L.P.; Dekker, C. *Nature* **2001**, *412*, 617.

(18) Slonczewski, J.C.; Phys. Rev. B **1989**, *39*, 6995; Moodera, J.S.; Nowak, J.; Kinder, L.R.; Tedrow, P.M. Phys. Rev. Lett. **1999**, *83*, 3029.

(19) Kirchner, S.; Zhu, L.; Si, Q.; Natelson, D. Preprint, *cond-mat/0507215*.

SYNOPSIS TOC (Word Style "SN_Synopsis_TOC"). If you are submitting your paper to a journal that requires a synopsis graphic and/or synopsis paragraph, see the Guide, Notes, Notice, or Instructions for Authors that appear in each publication's first issue of the year and the journal's homepage for a description of what needs to be provided and for the size requirements of the artwork.

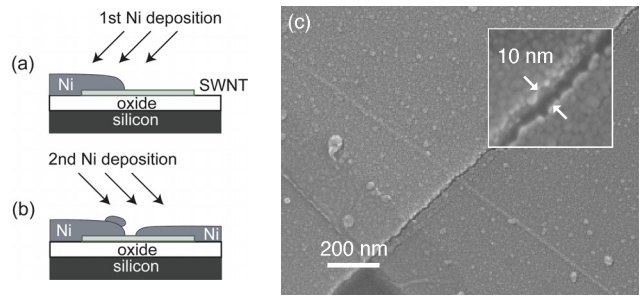

Figure 1. Fabrication of a short-channel nanotube device: **(a)** Deposition of the left ferromagnetic contact is performed first followed by, **(b)** deposition of the right ferromagnetic contact. The sample is angled so that the deposition of the right contact is partially shadowed by the left contact. **(c)** Field emission SEM image of a finished device. The inset shows a close-up of the contact separation, which is approximately 10 nm.

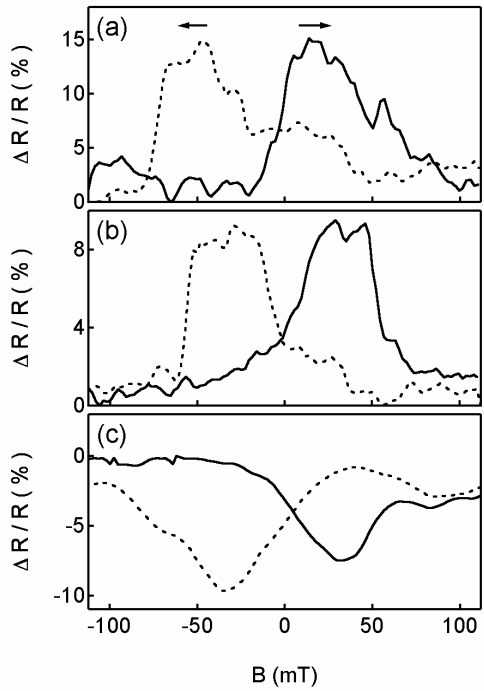

Figure 2. Percent change in resistance as a function of magnetic field for three nickel contacted short-channel SWNT devices. The solid (dashed) line corresponds to the positive (negative) sweep direction. Measurements were made at 4.2K with the gate grounded.

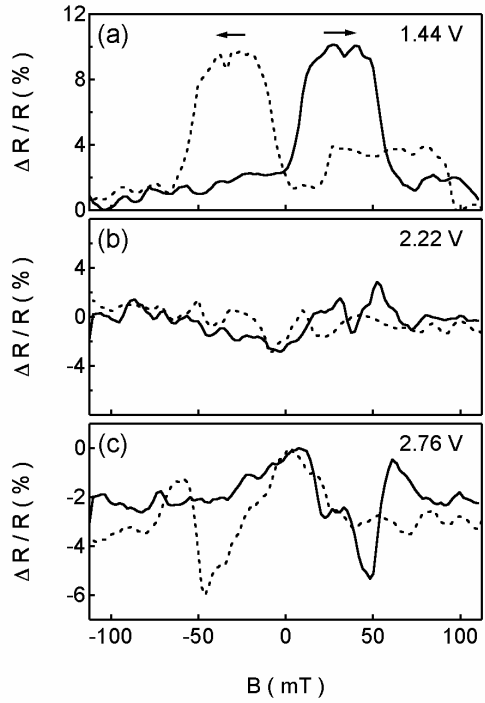

Figure 3. Magnetoresistance ratio of a nickel contacted, gated, short-channel SWNT device for three different values of the gate bias. The solid (dashed) line corresponds to the positive (negative) sweep direction. Measurements were made at 4.2K

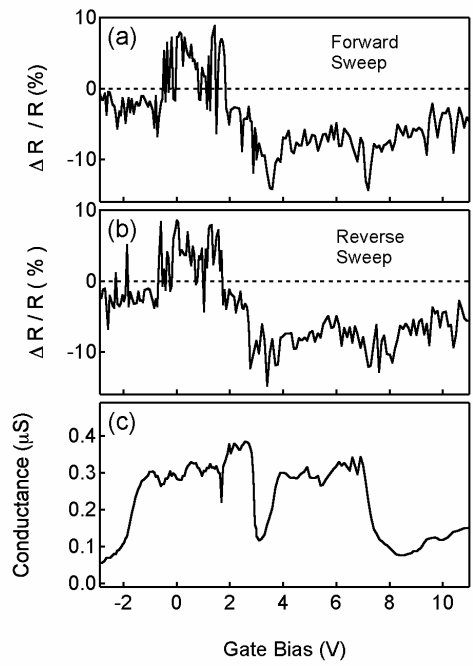

Figure 4. Magnetoresistance ratio of the device in Fig. 3 as a function of gate bias in the **(a)** forward and **(b)** reverse sweep directions. **(c)** Two-terminal conductance of the same device as a function of gate bias. All measurements are at 4.2 K.

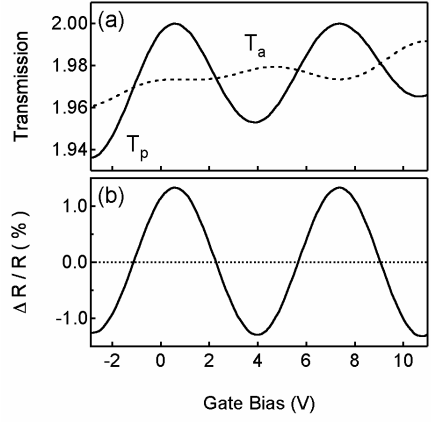

Figure 5. (a) Parallel and anti-parallel transmission coefficients and (b) the resistance ratio as a function of the gate voltage calculated using a one-dimensional model for a ferromagnetically contacted armchair nanotube. In the nickel contacts, the energy splitting between the spin-up and spin-down electrons $E = 1.0$ eV, the Fermi energy $E_F = 2.2$ eV and the effective mass $m^* = 1m_e$. For the nanotube, the length $L=10$ nm, $m^* = 1m_e$, and $k = k_0 + k$ with $k_0 = 0.8509/\text{Å}$, $k = E/hv_F$, and $v_F = 8.2 \times 10^5 \, m/s$. Here $E = \alpha e V_g$ with the gate efficiency factor $\alpha$ being taken to be 0.025.